\documentclass[
preprint,
showpacs,
nofootinbib,
nobibnotes,
amsmath,amssymb,
aps,
pra,
floatfix,
]{revtex4-2}

\usepackage{graphicx}
\usepackage{dcolumn}
\usepackage{bm}
\usepackage[colorlinks,linkcolor=blue,urlcolor=blue,
   citecolor=blue,]{hyperref}
\usepackage{multirow,wasysym}
\usepackage{rotating}
\usepackage{braket,mathtools}
\usepackage[	
margin=0.9in,
]{geometry}
\usepackage{natbib}
\usepackage[subrefformat=parens,labelformat=parens,caption=false]{subfig}
\usepackage{bibentry}

\begin{document}

\title{
Two-center basis generator method calculations for Li$^{3+}$, C$^{3+}$ and O$^{3+}$ ion impact
on ground state hydrogen
}
\author{Anthony C. K. \surname{Leung}}
\author{Tom \surname{Kirchner}}%
 \email{tomk@yorku.ca}

\affiliation{
 Department of Physics and Astronomy, York University, Toronto, Ontario,
 M3J 1P3, Canada
}%

\date{\today}

\begin{abstract}
The two-center basis generator method is used to obtain cross sections
for excitation, capture, and ionization in Li$^{3+}$, C$^{3+}$,
and O$^{3+}$ collisions with ground-state hydrogen at projectile energies from 1 to 100 keV/u.
The interaction of the C$^{3+}$ and O$^{3+}$ projectiles with the active electron is represented
by a model potential. Comparisons of cross sections with previously reported 
data show an overall good agreement, while discrepancies in capture
for C$^{3+}$ collisions at low energies are noted.
The present results show that excitation and ionization
are similar across the three collision systems, which indicates
that these cross sections are mostly dependent on the net charge
of the projectile only. The situation is different for the capture channel. 
\end{abstract}

\maketitle

%
%
\section{Introduction}

Collisions between partially stripped ions and neutrals (atoms or molecules)
are more commonly found in nature than collisions with bare ions. 
Partially stripped ion collisions have been a subject of interest in astrophysical~\cite{Cravens2000,Abu-Haija2007} and plasma applications~\cite{Bruhns2008}, and~thus, 
interest in accurate cross sections for electronic processes in these collisions remains high.
In recent times, the~International Nuclear Data Committee within the
International Atomic Energy Agency has expressed interest in cross sections from
collisions between bare or partially stripped projectiles with atomic hydrogen,
which are necessary for neutral beam modeling in fusion plasma~\cite{Chung2017}.

In this study, new calculated cross sections for collision systems involving ground-state hydrogen 
and ions of net charge $Q=3$ are reported.
Specifically, the~projectile ions Li$^{3+}$, C$^{3+}$ and O$^{3+}$ were chosen for this analysis. 
Cross sections for electron excitation, capture, and~ionization are compared
with the data that are available in the literature.
Currently, there is a broad coverage of cross sections for Li$^{3+}$-H(1s) collisions 
from the low- to the high-energy regimes (e.g., References~\cite{Shah1978, Fritsch1982, Murakami2008}).
Although data exist for C$^{3+}$ and O$^{3+}$ collisions with atomic hydrogen~\cite{Janev1988, Errea1991, Guevara2011}, there are some gaps 
for excitation and ionization data in the intermediate energy regime.
Therefore, the~objectives of this study are to 
report cross sections in these gaps, perform validity checks on existing ones, and~provide a comparison for the three ions to determine to
what  extent the net charge alone determines the cross~sections.

The approach used in the present theoretical analysis is the semiclassical, 
nonperturbative two-center basis generator method (TC-BGM) \cite{Zapukhlyak2005}.
As a close-coupling approach, the~main feature of the TC-BGM is its 
 dynamical basis that is adapted to the problem at hand, which provides
a practical advantage in terms of reaching convergence with smaller basis sets than 
required in standard approaches.
It is relevant to this study that the~TC-BGM was previously used   in References~\cite{Ludde2020a,Ludde2020} to obtain accurate capture and ionization 
cross sections for Li$^{3+}$-H(1s) collisions at impact energies of 10 keV/u and higher.
The present work focuses on collisions at impact energies from 1 to 100 keV/u. 

The article is organized as follows:
In Section~\ref{sec:theory}, an~overview of the TC-BGM for ion--atom
collisions is given. 
In Section~\ref{sec:results}, cross sections for the three collision systems are presented and discussed.
Finally, in Section~\ref{sec:conclusion}, concluding remarks are provided.
Atomic units $(\hbar=e=m_{e}=4\pi\epsilon_{0}=1)$ are used throughout
the article unless stated~otherwise.


%
%
\section{Theoretical Method \label{sec:theory}}
The focus of this study is on collisions in the low- and intermediate-impact
energy regimes, specifically from 1 to 100 keV/u.
The collisional framework used here
is based on the impact-parameter model within the semiclassical
approximation. 
Such a framework   has been used with the TC-BGM to solve the time-dependent
Schr\"{o}dinger equation (TDSE)
in previous studies and has also been described at  {some} length in a
prior work related to collisions with atomic hydrogen~\cite{Leung2019a}. 
For this reason, only a summary highlighting
the core ideas and some details regarding the potentials used are~given.

In the laboratory frame, the~hydrogen target is assumed to be fixed in space 
and the projectile ion travels in a straight-line path at constant speed $v_{\text{P}}$,
described by \mbox{$\mathbf{R}(t)=(b,0,v_{\text{P}}t)$}, where $b$ is the
impact parameter. The~objective is 
to solve the TDSE
for the initially occupied ground state in the target,
\begin{equation}\label{eq:TDSEs}
	i{\partial\over\partial t}\psi(\mathbf{r},t)=\hat{h}(t)\psi(\mathbf{r},t).
\end{equation}

For the present study, the~projectile ions under consideration are Li$^{3+}$, C$^{3+}$, and~O$^{3+}$.
This study assumes that the strongly bound electrons in the C$^{3+}$ and O$^{3+}$ ions remain
frozen during the collision. This allows the single-particle Hamiltonian to be decomposed~as
\begin{equation}
	\hat{h}(t) = -{1\over 2}\nabla^{2} + V_{\text{T}}(|\mathbf{r}|) + V_{\text{P}}(|\mathbf{r}_{\text{P}}|,t),
\end{equation}
where $\mathbf{r}_{\text{P}}$ is the electron position vector with respect to the projectile
and is related to that with respect to the target by $\mathbf{r}_{\text{P}}(t)=\mathbf{r}-\mathbf{R}(t)$.
The projectile Li$^{3+}$ is represented by the Coulomb potential with $Z_{\text{P}}=3$. For~the 
C$^{3+}$ projectile, two sets of cross sections are produced, where one is generated using
the optimized potential model (OPM) \cite{Engel1993} and another one using 
{the two-parameter analytic potential model of 
} Green, Sellin, and~Zachor (GSZ)~\cite{Green1969}.
The potential for the O$^{3+}$ projectile is represented with the GSZ model.

{The GSZ model can be viewed as a modified Hartree--Fock procedure
in which the expectation value of the many-electron Hamiltonian
is minimized with respect to a determinantal wave function 
constructed from orbitals which satisfy a one-electron
Schr\"odinger equation with the GSZ potential. The~minimization
fixes the two parameters involved. For~the calculations of this work,
we use the parameters reported in Ref. \cite{Szydlik1974}
for the C$^{2+}$(1s$^{2}$ 2s$^{2}$)
and the O$^{2+}$(1s$^{2}$ 2s$^{2}$ 2p$^{2}$) configurations such that the 
GSZ potentials exhibit the correct asymptotic behavior from the
viewpoint of an additional electron. One can criticize this
procedure for erroneously assuming that electron transfer to
the projectile produces a doubly charged ground-state ion, while
in reality electron capture mostly populates excited projectile
states (except at very high impact energies). 
However, it turns out that the orbital energy 
eigenvalues obtained from solving the one-electron Schr\"odinger 
equation with the GSZ potential are reasonably close to 
experimentally determined spectroscopic data \cite{Kramida2021}. 
In the case of carbon, 
for example, they do not deviate more than 5\% for the $n=3$ shell,
which is the dominant capture shell for most of the impact energy range
studied here. 
In addition, no assumption on the configuration in the projectile after
electron capture is made when generating the OPM potential and the 
corresponding bound states, which we use as an alternative.
In this case, we start with a fully numerical self-consistent potential for
the C$^{3+}$(1s$^{2}$2s) ion and subtract the exchange part in order
to represent the situation for an additional electron which does not contribute
to the charge distribution on the projectile.
This procedure is explained in Ref. \cite{Schenk2015}. 
The results obtained from both variants indicate that relatively 
small changes in the potential
model do not cause major discrepancies in the cross sections (see Sec.~\ref{sec:carbon})}.

The TDSE \eqref{eq:TDSEs} is solved 
by projection onto a finite set of basis states and
propagation using the TC-BGM.
The basis sets that were used for this analysis include: all $nlm$ states from $n=1$ to $n=5$ of
the hydrogen target, states from $n=1$ to $n=6$ of Li$^{2+}$, states from $n=2$ to $n=6$ of
C$^{2+}$ and O$^{2+}$ ions, and~45 BGM   {pseudostates} to account
for intermediate quasimolecular couplings and ionization to the continuum.
Details on the construction of the pseudostates and the calculation
of matrix elements can be found in Ref. 
\cite{Ludde2018}. 
Bound-state probabilities for finding the
electron on the target $p^{\text{tar}}$ or on the
projectile $p^{\text{cap}}$ are calculated from summing up the
transition probabilities within the bound-state basis sets, and~probabilities for total ionization $p^{\text{ion}}$ are obtained
from the unitarity criterion
\begin{equation}\label{eq:ion}
p^{\text{ion}}= 1- p^{\text{tar}}-p^{\text{cap}}.
\end{equation}

Finally, cross sections for the electronic transitions 
are obtained by integrating the probabilities over the impact
parameter
\begin{equation}\label{eq:cross}
\sigma = 2\pi\int_{0}^{b_{\text{max}}}bp(b) db,
\end{equation}
where $b_{\text{max}}$ (in a.u.) is the upper bound at which
the integral is cut in practice.
For the collision calculations reported here, an~upper bound of $b_{\text{max}}=20$ a.u. 
was more than sufficient to capture the asymptotic profile of these
transition probabilities.
It should also be noted that the conservation of unitarity \eqref{eq:ion}
was monitored in the present analysis  {by summing up the transition probabilities
to all BGM pseudostates after orthogonalizing them to the
bound-state basis sets. It} was found that 
deviations produced by the calculations are typically no larger than 1\%.

\section{Results and Discussion\label{sec:results}}

\subsection{Li$^{3+}$-H(1s)\label{sec:lithium}}

\begin{figure}[htbp]
\centering
\subfloat{\label{fig:li-ex}
\includegraphics[scale=0.83]{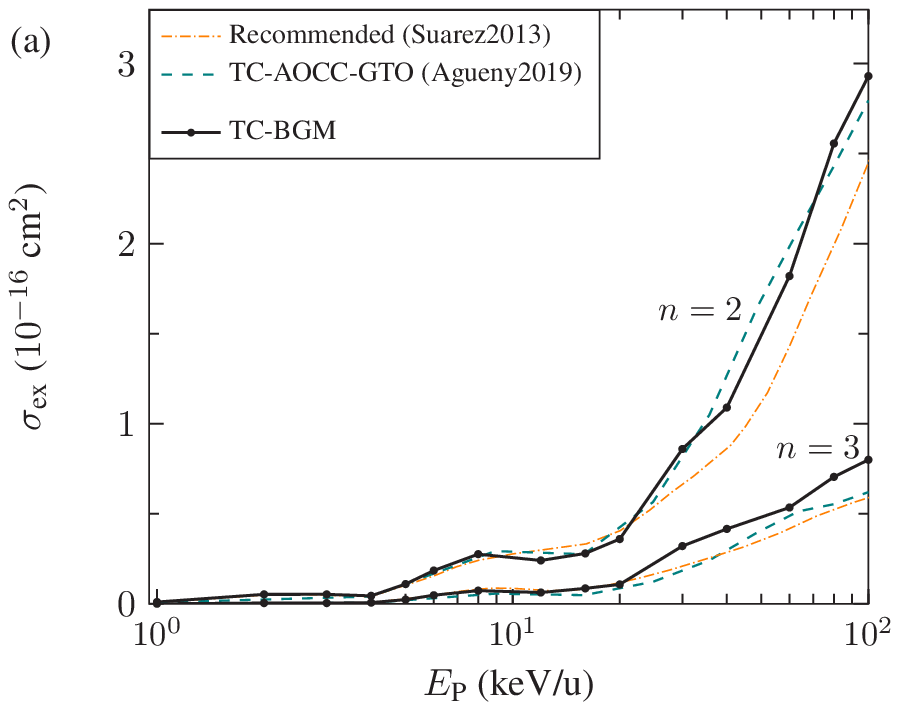}
}\\
\subfloat{\label{fig:li-cap}
\includegraphics[scale=0.83]{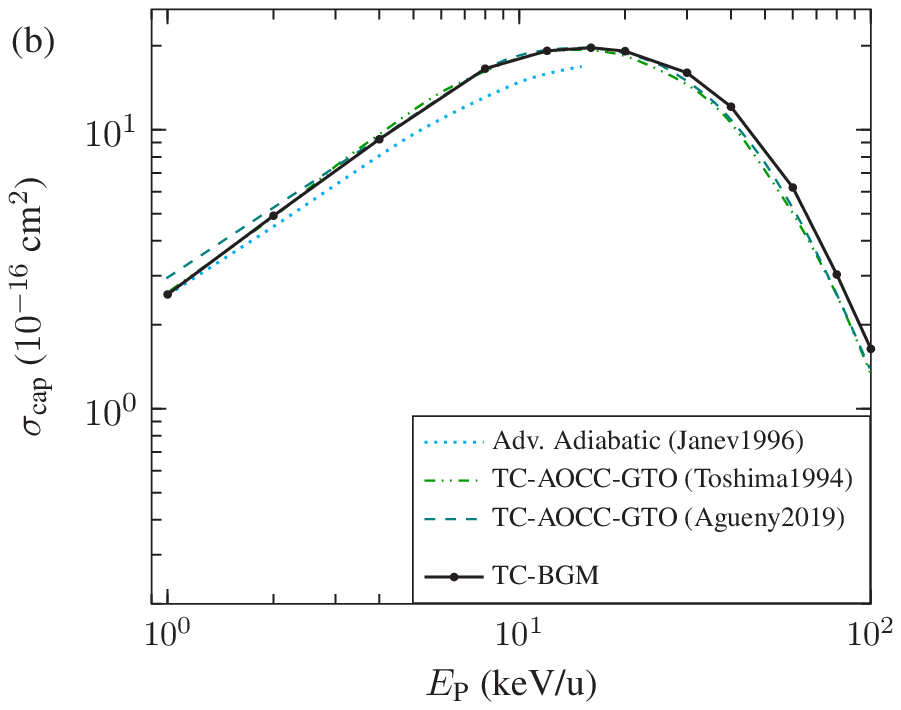}
}\\
\subfloat{\label{fig:li-ion}
\includegraphics[scale=0.83]{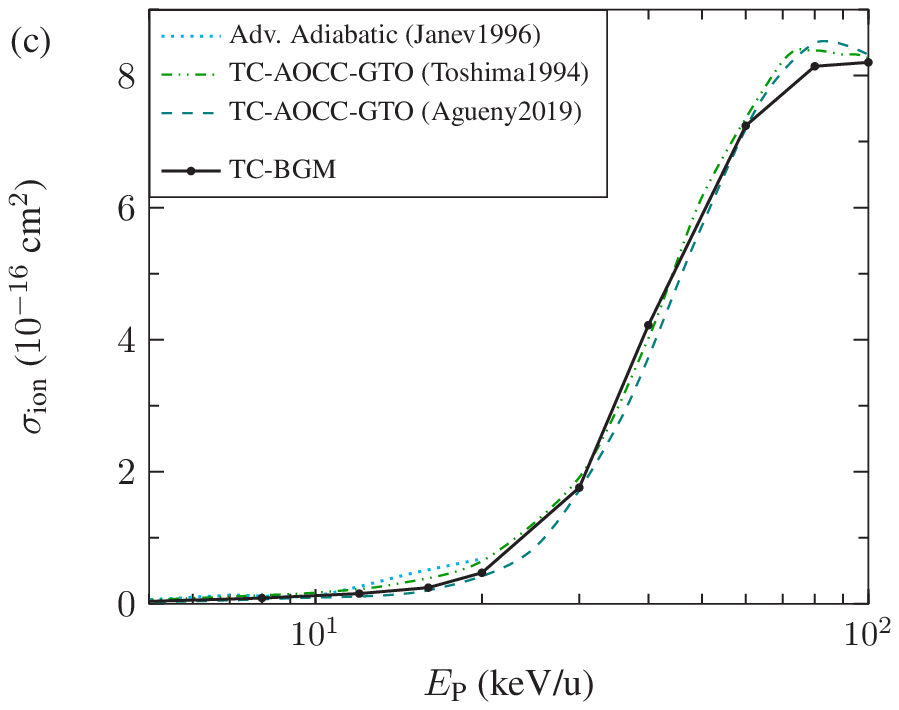}
}
\caption{Cross sections of (a) target-excitation, (b) electron capture, and
(c) ionization for Li$^{3+}$-H(1s) collisions from 1 to 100 keV/u.
Recommended data from Ref. \cite{Suarez2013}. Theory: present TC-BGM, 
TC-AOCC-GTO \cite{Toshima1994, Agueny2019a}, 
and Advanced Adiabatic \cite{Janev1996}.
}
\label{fig:li-h}

\end{figure}

The cross section results for the Li$^{3+}$-H(1s) collision system are shown
in Figure~\ref{fig:li-h}. Starting with the $n$-state excitation cross sections (Figure \ref{fig:li-ex}),
the present TC-BGM results are compared with other calculations.
The work by \citet{Suarez2013} reported recommended excitation cross sections
based on the two-center one-electron
diatomic molecule expansion covering 1 to 80 keV/u and the one-center Bessel expansion
that covers energies greater than 80 keV/u. 
The work by \citet{Agueny2019a} was based
on the two-center atomic orbital close-coupling method with Gaussian-type orbitals (TC-AOCC-GTO). 
Only the excitation of the $n=2$ and $n=3$ shells are presented, since transitions to higher
energy states are not important at these impact energies~\cite{Suarez2013}.
It is shown that the present TC-BGM results are in satisfactory agreement with 
previously reported cross sections~\cite{Suarez2013, Agueny2019a}.

In Figure~\ref{fig:li-cap}, total electron capture cross sections for the
Li$^{3+}$-H(1s) collision system are shown.
The present TC-BGM capture results show an impact-energy dependence, 
which is consistent with previously reported results,
namely, the~TC-AOCC-GTO~\cite{Agueny2019a}, another 
TC-AOCC-GTO calculation by \citet{Toshima1994}, and~the advanced
adiabatic method by \citet{Janev1996}.
Quantitatively, the~present cross sections are closest to the TC-AOCC-GTO results
~\cite{Toshima1994, Agueny2019a} at all impact energies. 
There are some differences in the advanced adiabatic calculations~\cite{Janev1996}
compared with the present results
but they are no more than 20\%, and~smaller at low energies
where the advanced  adiabatic method~\cite{Janev1996} is expected
to work~best.

{Figure~\ref{fig:li-ion} shows the ionization results for 
Li$^{3+}$-H(1s) collisions. The~present cross sections are compared with
those from the advanced adiabatic~\cite{Janev1996} 
and TC-AOCC-GTO~\cite{Toshima1994, Agueny2019a}} calculations.
It can be seen that the present TC-BGM results are consistent
with the previous calculations and show that ionization is important
in the intermediate energy regime at 10 keV/u and above but
negligible at lower energies.
They are also consistent with the aforementioned studies~\cite{Ludde2020a, Ludde2020}
using the TC-BGM on Li$^{3+}$-H(1s) collisions at 10 keV/u and above.
Overall, the~cross sections produced by the TC-BGM are in good agreement
with previous~calculations.

\subsection{C$^{3+}$-H(1s)\label{sec:carbon}}

\begin{figure}[htbp]
\centering
\subfloat{\label{fig:c3-ex}
\includegraphics[scale=0.83]{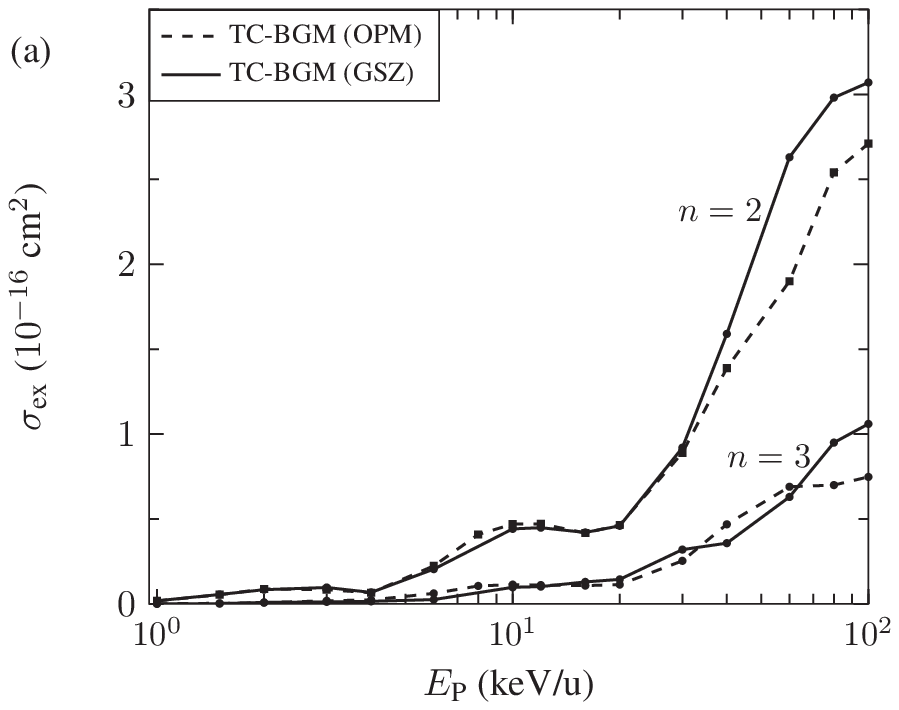}
}\\
\subfloat{\label{fig:c3-cap}
\includegraphics[scale=0.83]{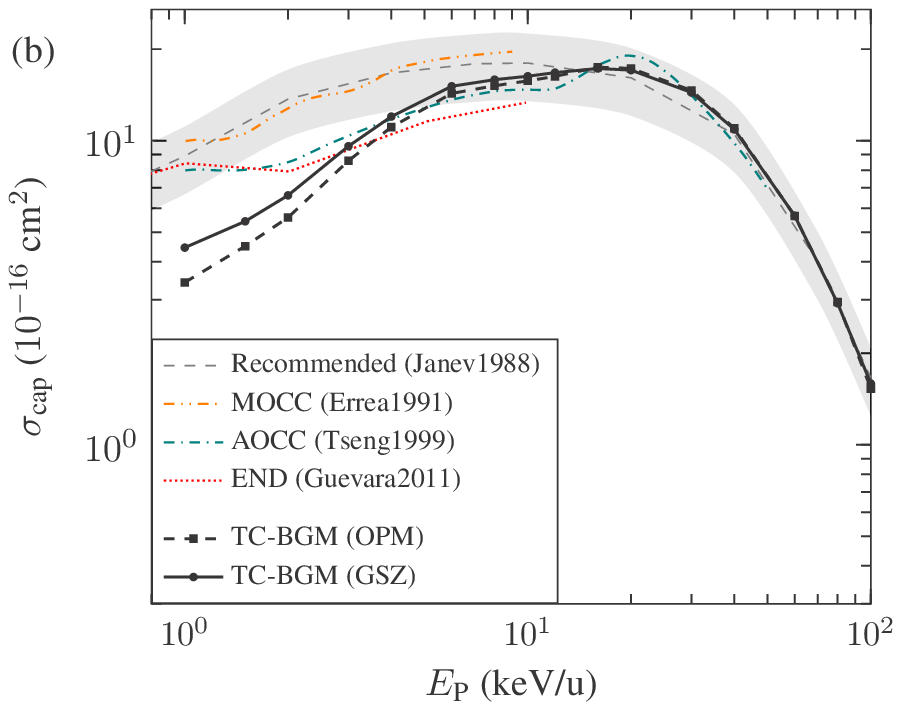}
}\\
\subfloat{\label{fig:c3-ion}
\includegraphics[scale=0.83]{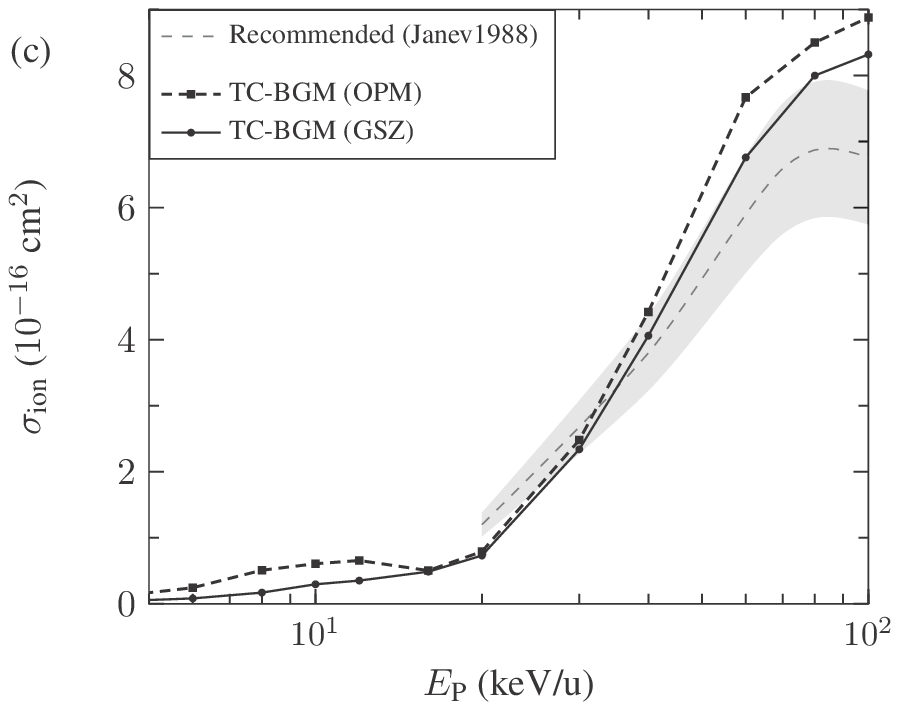}
}
\caption{Cross sections for C$^{3+}$-H(1s) collisions from 1 to 100 keV/u: (a) target-excitation;
(b) electron capture; (c) ionization. Recommended data from Ref. \cite{Janev1988}.
Theory: present TC-BGM,
MOCC \cite{Errea1991}, AOCC \cite{Tseng1999}, and END \cite{Guevara2011}.
}
\label{fig:c3-h}
\end{figure}

Shown in Fig.~\ref{fig:c3-h} are the cross section results for C$^{3+}$ collisions
with ground-state hydrogen.
The $n$-state excitation cross sections in Fig.~\ref{fig:c3-ex}
only include results from the present TC-BGM calculations since, to~our knowledge, no
other results are available in the literature. 
Two sets of results from TC-BGM calculations are shown,
where one set is based on using the OPM to represent the partially stripped ion and the other 
set is based on the GSZ potential. 
The excitation cross sections show the 
typical increasing behavior as the impact energy increases.
There are some discrepancies between the OPM and GSZ results around 100 keV/u but 
these differences decrease towards lower energies.
Similar to Li$^{3+}$ collisions, the~dominant excitation channel
in the C$^{3+}$-H(1s) system is $n=2$ followed by $n=3$. 
One can also see the qualitative and quantitative
similarities in the cross section profiles of these two systems 
with excitation processes being important at 20 keV/u and higher.
This observation reflects the fact that the target electron mainly
experiences an overall net charge of $Q=3$ (i.e., the same as for Li$^{3+}$ impact) and that
non-Coulombic interactions are of minor~importance.

In Fig.~\ref{fig:c3-cap}, the total capture cross sections for C$^{3+}$-H(1s)
collisions are presented. 
Again, two sets of TC-BGM results are shown where one set is
based on using the OPM for the projectile while the other one is based
on the GSZ potential.
Shown alongside   the present cross sections are
previously reported values from calculations using
a molecular-orbital close-coupling (MOCC) scheme~\cite{Errea1991}, 
calculations using an AOCC expansion~\cite{Tseng1999}, 
calculations based on the electron nuclear dynamics (END) approach~\cite{Guevara2011}, 
and a set of recommended values based on theoretical and experimental works 
compiled by \citet{Janev1988} with a 25\% uncertainty band.
Both the MOCC calculation~\cite{Errea1991} and the AOCC calculation~\cite{Janev1988}
are explicit two-electron calculations. 
One can see that the TC-BGM cross sections at low energies are significantly lower than
the previously reported results. 
Between the two sets of the present TC-BGM results, the capture cross section 
produced from using the GSZ potential is slightly closer to previous results than
those that obtained from using the OPM.
The discrepancies at low energies could be due to the present treatment 
of the C$^{3+}$ projectile, which assumes that the
screening of the nucleus is frozen throughout the course of the collision.
Interestingly, the~differences in capture between the OPM and GSZ results decrease as impact
energy increases; the opposite tendency is seen for excitation (Fig. \ref{fig:c3-ex}).

The overall agreement of the present TC-BGM ionization cross sections with the
recommended values~\cite{Janev1988} shown in Fig.~\ref{fig:c3-ion}
is satisfactory.
The cross section profile from the present calculations shows the expected
behavior where ionization is not important at 10 keV/u and below.
At higher energies, one can see the typical profile of an increasing cross
section. One can also see the quantitative similarities of Li$^{3+}$ and C$^{3+}$ impact 
between 10 and 100 keV/u.
Overall, it appears that the present cross section 
calculated using the GSZ potential is closer to the
recommended values than the results that are based on the~OPM.

\begin{figure}[htbp]
\includegraphics[scale=1]{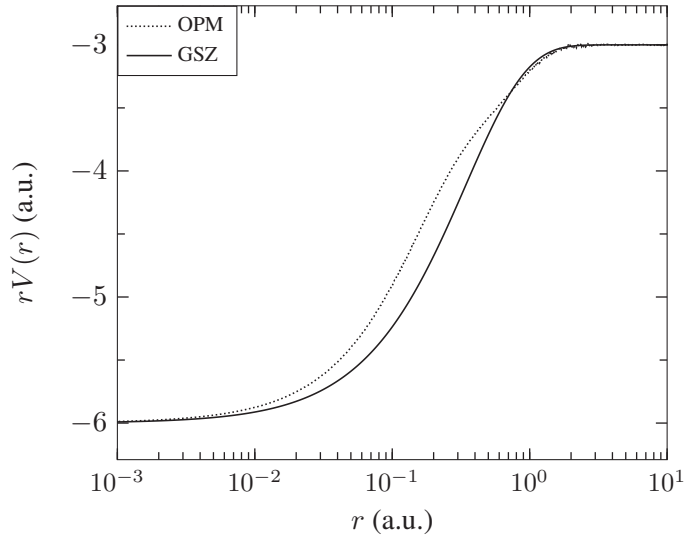}
\caption{Distance-weighted effective potential of the C$^{3+}$ projectile as a function
	of radial distance. Potentials calculated using the OPM \cite{Engel1993} 
	and GSZ \cite{Szydlik1974} are displayed.}
\label{fig:c3-pot}
\end{figure}

Figure~\ref{fig:c3-pot} illustrates the differences between the OPM and the GSZ potentials, where 
the $r$-weighted potentials are plotted with respect to the radial distance.
The potential profiles show identical asymptotic behaviors at short and long distances but display 
some differences in the $r\in[0.01, 1]$ a.u. interval where the GSZ potential
is mostly lower than the OPM.
In other words, the~GSZ models a potential that is more attractive 
than the~OPM.

\subsection{O$^{3+}$-H(1s)\label{sec:oxygen}}

\begin{figure}[htbp]
\centering
\subfloat{\label{fig:o3-ex}
\includegraphics[scale=0.83]{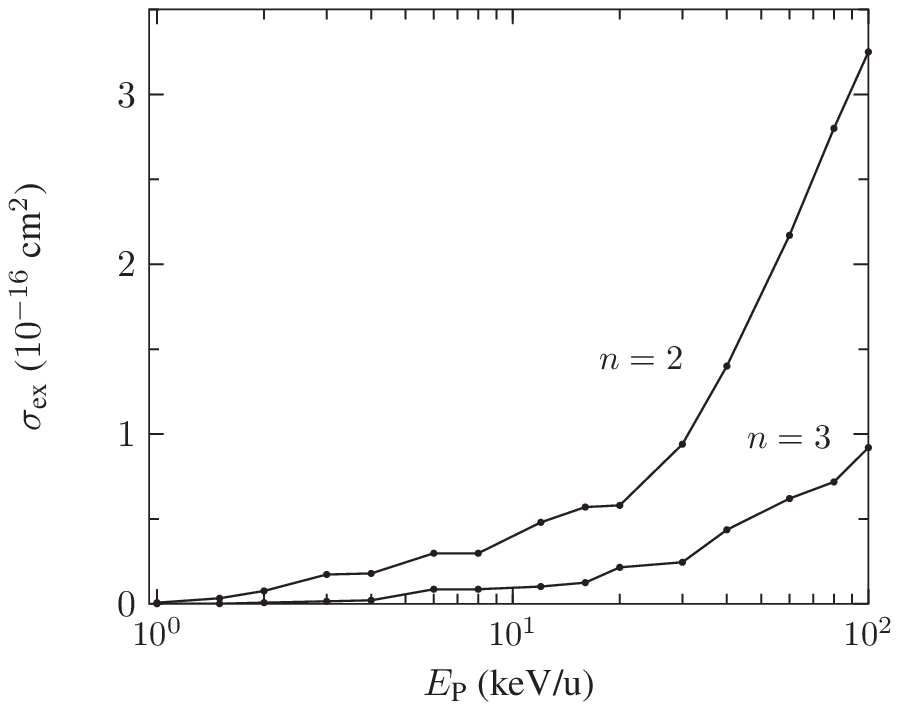}
}\\
\subfloat{\label{fig:o3-cap}
\includegraphics[scale=0.83]{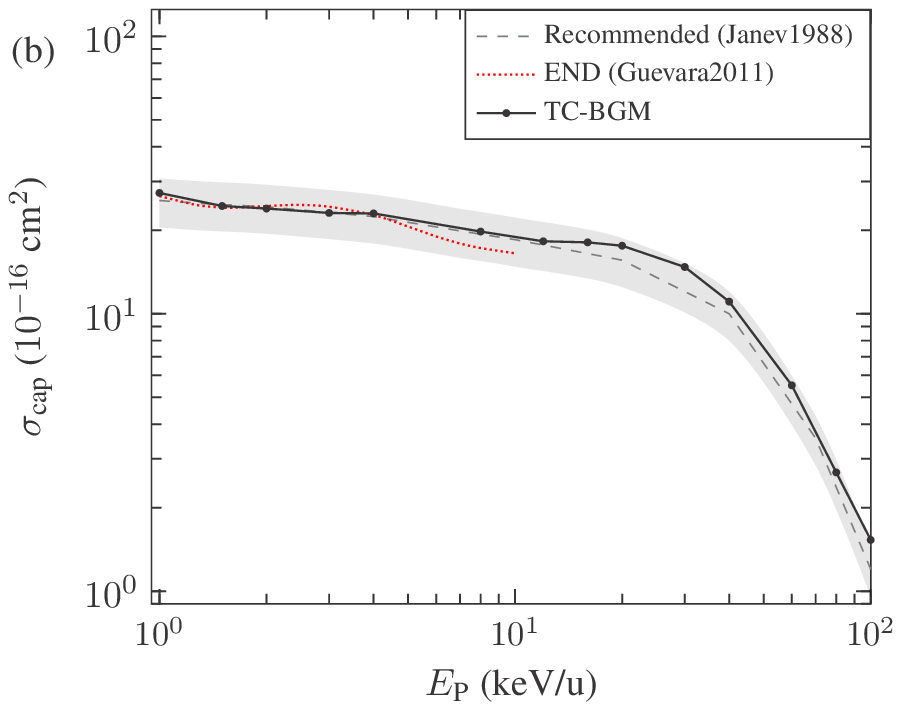}
}\\
\subfloat{\label{fig:o3-ion}
\includegraphics[scale=0.83]{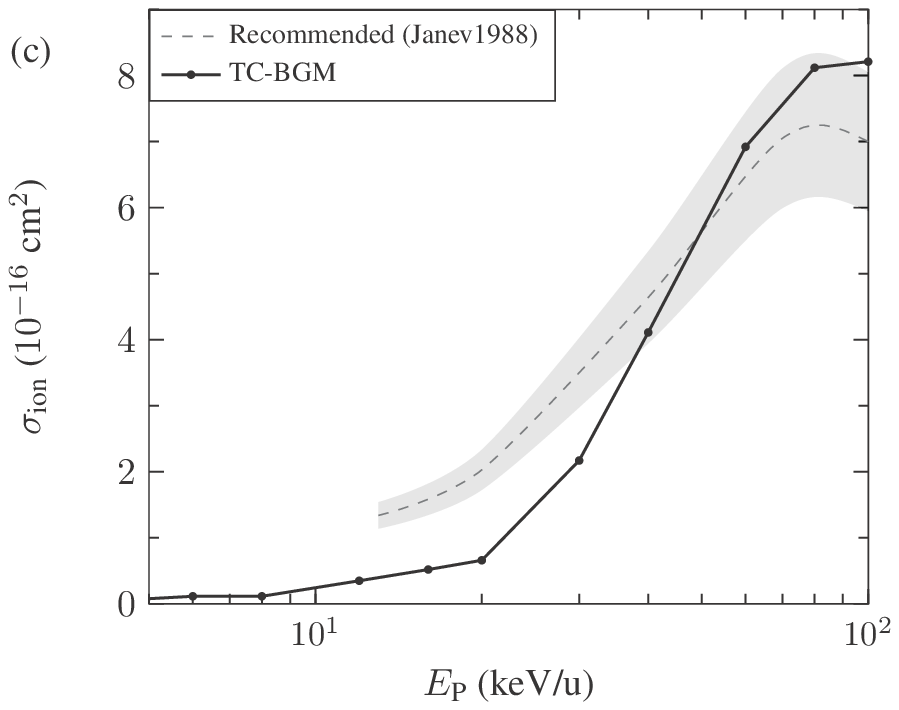}
}
\caption{Cross sections for O$^{3+}$-H collisions from 1 to 100 keV/u: (a) target-excitation;
(b) electron capture; (c) ionization.
Recommended data from Ref. \cite{Janev1988}.
Theory: present TC-BGM and
END \cite{Guevara2011}.
}
\label{fig:o3-h}
\end{figure}

Figure~\ref{fig:o3-h} shows the set of cross sections 
for O$^{3+}$-H(1s) collisions. 
The excitation cross sections in Fig.~\ref{fig:o3-ex}
show the typical increasing behavior between 1 and 100 keV/u.
Moreover, one can again see that the dominant channel is $n=2$
and the quantitative similarities to the results for Li$^{3+}$
and C$^{3+}$ collisions. Specifically, a~cross section value 
of approximately $3\times 10^{-16}$ cm$^{2}$ for the
$n=2$ channel is obtained at 100 keV/u across
all three collision~systems.

In Fig.~\ref{fig:o3-cap}, the total capture cross sections calculated from the present
TC-BGM are compared with the previously reported recommended values~\cite{Janev1988}
and the more recent results from END calculations~\cite{Guevara2011}. 
The present results are quantitatively consistent with the previous results
and are within the uncertainty range of the recommended values~\cite{Janev1988}. 
This is in contrast to the observation that was made for the capture
cross section in C$^{3+}$ collisions (Fig. \ref{fig:c3-cap}), where discrepancies 
with the END and with other calculations are significant at low impact energies.
In addition, the~capture profile here is noticeably 
different from that of the Li$^{3+}$ (Fig. \ref{fig:li-cap})
and C$^{3+}$ (Fig. \ref{fig:c3-cap}) collisions, with~capture by O$^{3+}$ ions 
showing a monotonic decrease over the entire energy range~displayed.

Figure \ref{fig:o3-ion} shows the total
ionization cross section.
The present TC-BGM results are compared with the previously compiled recommended
values \cite{Janev1988}. One can see that the present results
are well within the uncertainty range of the recommended values
between 40 and 100 keV/u but fall short at lower energies.
Given that only these recommended values are available, additional
independent studies are necessary to provide further validation
of the present results.

\section{Conclusions\label{sec:conclusion}}

In this work, we reported TC-BGM cross section calculations for 
Li$^{3+}$, C$^{3+}$, and O$^{3+}$ collisions with ground-state
hydrogen from 1 to 100 keV/u.
Cross sections for electron excitation, capture, and ionization
were obtained for all three systems.
Overall, we found satisfactory agreement between 
the present cross sections and previously reported values.

\begin{figure}[htbp]
\centering
\includegraphics[scale=1]{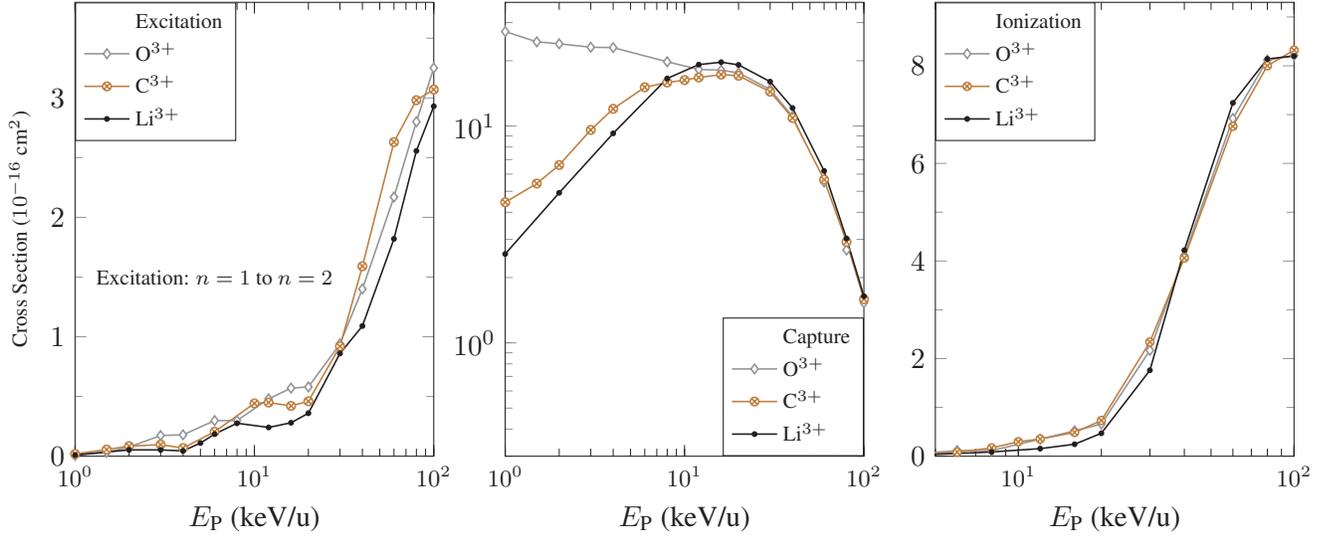}
\caption{TC-BGM cross sections of Li$^{3+}$-, C$^{3+}$-, and O$^{3+}$-H(1s) collisions 
plotted with respect to impact energies for
target-excitation from $n=1$ to $n=2$ (left panel), total electron capture
(middle panel), and total ionization (right panel). \label{fig:all}}
\end{figure}

We also drew a few comparisons of the cross sections across
the different projectiles. This is summarized in Fig. \ref{fig:all}
where total cross sections for excitation, capture, and ionization
for all three collision systems are plotted together.
One can observe the similarities in the energy dependence of the excitation and ionization
cross sections across the three systems. For capture, differences 
in cross sections are significant at 10 keV/u and below, 
indicating that the precise form of the screening of the projectile nucleus
matters.

\section*{Acknowledgments}
Financial support from the Natural Sciences and Engineering Research 
Council of Canada (NSERC) (RGPIN-2019-06305) is gratefully acknowledged.
This work was made possible with the high-performance computing 
resources provided by Compute/Calcul Canada.

\section*{Appendix}

The cross section data from the TC-BGM calculations for 
Li$^{3+}$, C$^{3+}$, and O$^{3+}$ collisions with 
ground-state hydrogen
are presented in Tables \ref{tab:ex-data}, \ref{tab:cap-data},
and \ref{tab:ion-data}. 
For the C$^{3+}$ and O$^{3+}$ projectiles the GSZ potential has been used.

\begin{table}[h]
\caption{\label{tab:ex-data} $n$-state selective 
excitation cross sections (10$^{-16}$ cm$^{2}$) 
for Li$^{3+}$, C$^{3+}$, and O$^{3+}$ collisions with 
ground-state hydrogen from
1 to 100 keV/u.}
\centering
\begin{ruledtabular}

\begin{tabular}{lccccccccr}
E(keV/u)         & \multicolumn{2}{c}{Li$^{3+}$-H(1s)} & \multicolumn{2}{c}{C$^{3+}$-H(1s)} & \multicolumn{2}{c}{O$^{3+}$-H(1s)} \\
         \cline{2-3} \cline{4-5} \cline{6-7}
 & n=2              & n=3              & n=2              & n=3             & n=2              & n=3             \\
\hline
1        & 0.010            & 0.001            & 0.019            & 0.001           & 0.007            & $<$0.001           \\
2        & 0.053            & 0.005            & 0.084            & 0.009           & 0.076            & 0.007           \\
3        & 0.053            & 0.005            & 0.097            & 0.012           & 0.173            & 0.015           \\
4        & 0.045            & 0.007            & 0.068            & 0.016           & 0.179            & 0.021           \\
6        & 0.185            & 0.048            & 0.205            & 0.026           & 0.298            & 0.086           \\
8        & 0.276            & 0.074            & 0.442            & 0.098           & 0.341            & 0.071           \\
12       & 0.241            & 0.063            & 0.449            & 0.102           & 0.480            & 0.102           \\
16       & 0.280            & 0.086            & 0.421            & 0.130           & 0.570            & 0.125           \\
20       & 0.360            & 0.108            & 0.460            & 0.145           & 0.580            & 0.215           \\
30       & 0.860            & 0.321            & 0.920            & 0.320           & 0.940            & 0.245           \\
40       & 1.090            & 0.416            & 1.590            & 0.358           & 1.400            & 0.436           \\
60       & 1.820            & 0.535            & 2.630            & 0.630           & 2.170            & 0.620           \\
80       & 2.556            & 0.705            & 2.981            & 0.950           & 2.800            & 0.718           \\
100      & 2.930            & 0.800            & 3.070            & 1.060           & 3.250            & 0.920          
\end{tabular}
\end{ruledtabular}
\end{table}

\newpage

\begin{table}[h]
\caption{\label{tab:cap-data}Total-capture cross sections (10$^{-16}$ cm$^{2}$) 
for Li$^{3+}$, C$^{3+}$, and O$^{3+}$ collisions with 
ground-state hydrogen from
1 to 100 keV/u.}
\begin{ruledtabular}
\begin{tabular}{lccr}
E(keV/u) & Li$^{3+}$-H(1s) & C$^{3+}$-H(1s) & O$^{3+}$-H(1s) \\
\hline
1        & 2.57            & 4.45           & 27.26          \\
2        & 4.92            & 6.61           & 23.95          \\
3        & 7.24            & 9.59           & 23.09          \\
4        & 9.25            & 12.00          & 23.00          \\
6        & 13.86           & 15.10          & 21.10          \\
8        & 16.57           & 15.87          & 19.79          \\
12       & 19.20           & 16.74          & 18.25          \\
16       & 19.70           & 17.27          & 18.09          \\
20       & 19.15           & 17.06          & 17.59          \\
30       & 16.02           & 14.40          & 14.74          \\
40       & 12.11           & 10.89          & 11.06          \\
60       & 6.21            & 5.66           & 5.52           \\
80       & 3.03            & 2.92           & 2.68           \\
100      & 1.64            & 1.59           & 1.53          
\end{tabular}
\end{ruledtabular}
\end{table}

\newpage

\begin{table}[h]
\caption{\label{tab:ion-data}Total-ionization cross sections (10$^{-16}$ cm$^{2}$) 
for Li$^{3+}$, C$^{3+}$, and O$^{3+}$ collisions with 
ground-state hydrogen from
1 to 100 keV/u.}
\begin{ruledtabular}
\begin{tabular}{lccr}
E(keV/u) & Li$^{3+}$-H(1s) & C$^{3+}$-H(1s) & O$^{3+}$-H(1s) \\
\hline
1        & $<0.001$        & 0.002          & 0.001          \\
2        & 0.003           & 0.011          & 0.007          \\
3        & 0.006           & 0.026          & 0.030          \\
4        & 0.014           & 0.028          & 0.034          \\
6        & 0.042           & 0.082          & 0.115          \\
8        & 0.087           & 0.172          & 0.115          \\
12       & 0.155           & 0.352          & 0.350          \\
16       & 0.245           & 0.487          & 0.502          \\
20       & 0.466           & 0.731          & 0.660          \\
30       & 1.760           & 2.340          & 2.173          \\
40       & 4.215           & 4.056          & 4.019          \\
60       & 7.236           & 6.757          & 6.811          \\
80       & 8.141           & 8.003          & 8.118          \\
100      & 8.198           & 8.320          & 8.209   
\end{tabular}
\end{ruledtabular}
\end{table}

\newpage
\bibliographystyle{apsrev4-2}

\bibliography{mainbib}

\end{document}